\documentstyle[emulateapj,psfig]{article}
\input epsf.tex

\begin{document}

\title{Precursors of gamma-ray bursts: a clue to the burster's nature}

\author{Maxim Lyutikov}

\affil{Canadian Institute for Theoretical Astrophysics, Toronto,
Ontario, M5S 3H8, Canada}

\author{Vladimir V.~Usov} 

\affil{Department of Condensed Matter Physics, Weizmann Institute, 
Rehovot 76100, Israel}

\date{Received   / Accepted  }

\begin{abstract}
In relativistic strongly magnetized winds outflowing from  the fast-rotating
compact  progenitors of gamma-ray bursts (GRBs) there are three regions 
where  powerful high-frequency emission may be generated: (i) the  thermal 
photosphere, (ii) the  region of the internal wind instability and  
(iii) the region of the wind  interaction with an ambient gas. This results
in a multicomponent structure of GRBs. The emission from the thermal 
photosphere may be observed as a weak precursor to the  main burst.
The precursor should have a blackbody-like spectrum with the mean energy
of photons of $\sim 1$~MeV, and its intensity should be tens to hundreds 
of times smaller than that of the main GRB emission. Observations 
of such precursors with future $\gamma$-ray missions like GLAST
can clarify the nature of bursters.


\end{abstract}


\section{Introduction}
Detections of absorption and emission features at high redshifts
in optical afterglows of GRBs and their host galaxies clearly 
demonstrate that the GRB sources lie at cosmological distances 
(e.g., Piran \cite{Piran99}; Vietri \cite{Vietri99}). Despite 
such a great advance, the nature of the GRB sources is still unknown. 
Several currently popular models posit as the energy-releasing event
coalescence of two neutron stars (Blinnikov et al. \cite{Blinnikov84}; 
Paczy\'nski \cite{Paczynski86}), 
the collapse of a massive star (Woosley \cite{Woosley93}; Paczy\'nski
\cite{Paczynski98}); or accretion 
induced collapse of magnetic white dwarfs to neutron stars 
(Usov \cite{Usov92}; Yi \& Blackman \cite{Yi97};
Ruderman, Tao \& Klu\'zniak \cite{Ruderman00}).
In all these models, fast-rotating compact objects like millisecond 
pulsars with the surface magnetic field of $\sim 10^{15}-10^{17}$ G
may be formed (Usov \cite{Usov92};
Blackman, Yi, \& Field \cite{Blackman96}; Katz \cite{Katz97}; 
Klu\'zniak \& Ruderman \cite{Kluzniak98}; Vietri \& Stella 
\cite{Vietri98}; Spruit \cite{Spruit99};  
Wheeler et al. \cite{Wheeler00}). The rotation of these objects 
decelerates on a time scale of seconds, and relativistic strongly 
magnetized winds (RSMWs) are generated (e.g., Usov 
\cite{Usov99}). These winds are plausible sources of cosmological GRBs.
In \S 2 we discuss radiation from RSMWs. More than 2500 GRBs have
been detected by BATSE. The large area detectors of BATSE provided 
excellent light curves of GRBs. Some properties of the light curves 
expected for GRBs in this model are considered and compared with 
available data in \S 3. Our main results are discussed in \S 4.

\section{Radiation from RSMW}

Our proposal applies more generally, but, for the sake of concreteness,
we consider millisecond pulsars with extremely strong magnetic fields
as the sources of RSMWs. The wind outflowing from a such a pulsar 
is Poynting flux$-$dominated, i.e., $\sigma = L_{\pm}/L_{_P}
\ll 1$, where $L_{_P}\simeq {B_{_S}^2R^6\Omega^4/ c^3}$
is the pulsar luminosity  in the Poynting flux, $ L_{\pm}$ 
is its luminosity in both electron-positron pairs and radiation,  
$R\simeq 10^6$ cm is the radius of the pulsar and $\Omega$ is 
its angular velocity. For typical parameters,  
$\Omega\simeq 10^4$~s$^{-1}$ and $B_{_S}\simeq 10^{16}$~G, we have
$L_{_P}\simeq 3\times 10^{52}$ erg ~s$^{-1}$ that is about the
luminosities of cosmological $\gamma$-ray bursters. The more plausible 
value of $\sigma$ is $\sim 0.01-0.1$ (Usov 1994). 

At $L_{_P}\sim 10^{52}$ erg  s$^{-1}$ and $\sigma \sim 0.01-0.1$, the 
density of electron-positron pairs near the compact object is so high
that the optical depth for radiation is as high as $\sim 10^{12}$.
In this case, the outflowing plasma and 
radiation are in thermodynamic equilibrium, and 
the outflowing hot wind may be completely described by $L_\pm$ 
and the mass-loss rate in baryons, $\dot M$. The values of 
$L_{\pm}$, $\sigma$ and $\dot M$ may be estimated from 
observational data on GRBs (see below). 

We identify three possible  types of high-frequency emission from RSMWs:
(i) thermal radiation from the wind photosphere,
(ii) non-thermal radiation generated  because of  the development of
internal wind instabilities
(iii)  non-thermal radiation produced by
 the interaction  between RSMWs  and  an ambient plasma
(see Fig. 1).

\subsection{Thermal radiation from the wind photosphere}
 
During outflow, the wind plasma accelerates and its density decreases. 
At the wind photosphere, $r\simeq r_{\rm ph}$, the 
optical depth for the bulk of thermal photons is $\sim 1$, and 
these photons propagate freely at $r> r_{\rm ph}$. For a spherical 
electron-positron wind with $\dot M=0$, the radius of 
the photosphere and the Lorentz factor of the outflowing plasma at this 
radius are  determined by the condition that  the comoving
plasma temperature at $r_{\rm ph}$ is $T_0\simeq 2\times 10^8$~K
and by the law $\Gamma \propto r$ for $ r \leq r_{\rm ph}$
(Paczy\'nski \cite{Paczynski90}; Usov \cite{Usov94}) :

\begin{equation}
r_{\rm ph} \simeq 3\times 10^8 \left({L_\pm\over 
10^{51} ~ {\rm erg ~ s}^{-1}}\right)^{1/4}\,{\rm cm}\,,
\end{equation}

\begin{equation}
\Gamma_{\rm ph}\simeq 10^2\left({L_\pm\over 10^{51}~{\rm erg~s}^{-1}}
\right)^{1/4}\,.
\end{equation}


The mean energy of observed photons is
$\langle\varepsilon_\gamma \rangle \simeq 2\Gamma_{\rm ph}T_{_0} \sim 1$~MeV.

For an optically thick
electron-positron wind, the thermal luminosity
of the wind photosphere, $L_{\rm ph}$, practically
coincides with $L_\pm$  (Usov \cite{Usov99}).

Presence of  baryons in RSMWs may change the above estimates.
If the mass-loss rate in baryons is small,  
 $\dot M < \dot M_1
\simeq 10^{-9}(L_\pm/10^{51}~{\rm erg~s}^{-1})^{3/4}~M_\odot~{\rm s}
^{-1}$, then the  baryonic loading of the wind is  negligible.
For larger mass-loss rates, 
$\dot M_1<\dot M <\dot M_2$, the following relations are valid,
$r_{\rm ph}\propto\dot M/
\Gamma^2_{\rm ph}$ and $\Gamma_{\rm ph} \propto r_{\rm ph}$, where  
$\dot M_2 = 0.5\times 10^{-6}(L_\pm /10^{51}~{\rm erg~s}^{-1})^{3/4}
~M_\odot~{\rm s}^{-1}$ (e.g., Paczy\'nsky \cite{Paczynski90}).
These relations and equations (1) and (2) yield
\begin{equation}
r_{\rm ph}\simeq 3\times 10^8\left({\dot M\over 10^{-9}~M_\odot~{\rm s}
^{-1}}\right)^{1/3}\,\,{\rm cm}\,,
\end{equation}

\begin{equation}
\Gamma_{\rm ph}\simeq 10^2 \left({\dot M\over 10^{-9}~M_\odot
~{\rm s}^{-1}}\right)^{1/3}\,.
\end{equation}

\noindent
In this case, the photospheric luminosity is about $L_\pm$, similar to 
a pure electron-positron wind, $(L_\pm - L_{\rm ph})/L_\pm = 
\dot Mc^2\Gamma_{\rm ph}/L_\pm\simeq (\dot M/M_2)^{4/3}$.
The baryon loading with $\dot M_1<\dot M <
\dot M_2$ does not change essentially the mean energy of photons 
radiated from the wind photosphere. This is because 
$\langle\varepsilon_\gamma\rangle$ is $\propto T_0\Gamma_{\rm ph}$ while 
$T_0\propto r_{\rm ph}^{-1}$ and $\Gamma_{\rm ph}\propto r_{\rm ph}$.

For even larger mass-loss rates, $\dot M >\dot M_2$, the baryon loading is very important
(Paczy\'nsky \cite{Paczynski90}):
\begin{equation}
r_{\rm ph}\simeq 3\times 10^9\left({\dot M\over \dot M_2}
\right)^3\left({L_\pm\over 10^{51} ~ {\rm erg~s}^{-1}}
\right)^{1/4}~{\rm cm}\,,
\end{equation}
\begin{equation}
\Gamma_{\rm ph}\simeq {L_\pm\over \dot Mc^2}\simeq 10^3
\left({\dot M\over \dot M_2}\right)^{-1}
\left({L_\pm\over 10^{51}~{\rm erg~s}^{-1}}\right)^{1/4}\,,
\end{equation}
Thus, for 
 $\dot M > \dot M_2$  the photospheric luminosity  and the mean 
energy of thermal photons decrease rapidly ($\propto \dot M^{-8/3}$)
 with  increasing $\dot M$.

\subsection{Non-thermal radiation from RSMWs}

The main source of energy for non-thermal radiation from RSMWs is
the magnetic field energy. RSMWs outflowing from the GRB progenitors
should resemble the winds from  rotation-powered neutron stars (pulsars)
(Coroniti \cite{Coroniti1990}; Melatos \& Melrose \cite{Melatos96}). They 
have two main components:   helical and  striped  magnetic fields; the
striped component alternates in polarity on a scale length of 
$\pi (c/\Omega )\sim 10^7$ cm. At small  distances  
the magnetic field of the wind is frozen into the plasma.
As the plasma flows out, the plasma density
decreases in proportion to $r^{-2}$ reaching a radius
$r_f\sim 10^{13}-10^{14}$~cm  where
it becomes less than the  critical charge density  
(Goldreich-Julian density) required for the magnetic field to be frozen.
At  $r>r_f$  
the plasma density is not sufficient to screen displacement currents
and  the striped component of the 
wind field is transformed into large-amplitude electromagnetic
waves (LAEMWs) due to the  development of magneto-parametric
instability (Usov \cite{Usov94}). The energy of  the LAEMWs is,
in turn, transfered almost completely to the 
electron-positron pairs 
and then to X-ray and $\gamma$-ray
photons
with the typical energy 
$\langle\varepsilon_\gamma\rangle\sim 1$ MeV (Usov \cite{Usov94};
Blackman et al. \cite{Blackman96}; Lyutikov \& Blackman
\cite{Lyutikov00}).



At $r\gg r_f$, the magnetic field is helical 
everywhere in the  outflowing wind.
This wind expands more or less freely up to the distance 
$r_{\rm{dec}}\sim 10^{17}$~cm at which 
deceleration of the wind because of its interaction with an interstellar
gas becomes important.
It was suggested by M\'{e}sz\'{a}ros and Rees (\cite{MR92}) 
that at this distance  an essential part of the wind 
energy may be radiated in X-rays and $\gamma$-rays. This suggestion 
was recently confirmed by numerical simulations (Smolsky 
\& Usov \cite{SU96}, \cite{SU00}; Usov \& Smolsky \cite{US98}). 
In these simulations it was shown that at $r\sim r_{\rm dec}$ in 
the process of the RSMW -- ambient gas interaction a shock-like radiating 
layer forms ahead of the wind front and about 20\% of the wind energy 
may be transferred to high-energy electrons of this layer
and then to high frequency (X-ray and $\gamma$-ray) emission.

\section{Light curves: theory meets observations}

The fact that in RSMWs there are  several
radiating regions  results in a multicomponent 
structure of GRBs. Since the radiating regions 
are at different distances from the wind source, sub-pulses generated 
at the different regions are shifted  in time with respect to each other. 
The first sub-pulse that may be observed in the light curve 
of a GRB is radiated from the wind photosphere. This sub-pulse looks
like a weak precursor of the bulk emission
(c.f. Hansen \& Lyutikov (\cite{Hansen00})). It is expected 
that the peak intensity of such a precursors is $\sim 10-10^2$ 
times smaller than that of the remaining emission. The precursor 
has a blackbody-like spectrum, and its duration is about
the characteristic time of deceleration of the neutron star rotation
$\tau_{_\Omega}\sim 0.1-10$~s. If the outflow from the
millisecond pulsar is non-stationary, the radiation of the precursor may 
be strongly variable. The time scale of this variability may be as small 
as $\Delta\tau_{\rm ph}=r_{\rm ph}/2\Gamma_{\rm ph}^2c\sim 10^{-6}$~s. 
This value is a few orders of magnitude less than the pulsar 
period $P=2\pi /\Omega\sim 10^{-3}$~s. If the angle $\vartheta$
between the rotational and magnetic axes of the pulsar is non-zero, 
$\vartheta \neq 0$, the density of the outflowing plasma may be 
modulated with the pulsar period. In this case the precursor 
radiation may be periodical with the pulsar 
period. Using equations (5) and (6) we can see that
this periodical variability may be observed 
even if the mass-loss rate in baryons $\dot M$ is as high as
$\sim 10^{-6}~M_\odot$~s$^{-1}$. Hence, observations of both
the spectra of GRB precursors and their short-time variability can 
test the GRB model. 

The second component that may be observed in the light curve of a GRB
is generated at the distance $r\simeq r_f\sim 10^{13}-10^{14}$~cm 
where the striped component
of the wind field is transformed into LAEMWs.
For its generation it is necessary that $\vartheta$ is non-zero.
This component has a non-thermal spectrum.
The time delay between the precursor and the second component 
is $\Delta\tau _f\simeq r_f/2c\Gamma_0^2$, where $\Gamma_0$ is the 
Lorentz factor of the outflowing plasma at $r\gg r_{\rm ph}$.
The value of $\Gamma_0$ is somewhere between $\Gamma_{\rm ph}$ and 
$\Gamma_f\sim 10^3$. Using this and equations (2), (4) and (6), 
for typical parameters, $L_\pm\simeq 10^{50}$~erg ~s$^{-1}$,
$B_{_S}\simeq 10^{16}$~G, $\Omega\simeq 10^4$~s$^{-1}$ and $\dot M <
10^{-6}M_\odot$~s$^{-1}$, the value of $\Delta\tau_f$ is
in the range from one millisecond  to a few seconds. 
In the extreme case when $\Delta\tau_f$ is as small as 
$\sim 10^{-3}$~s, the precursor and the second component practically
coincide and cannot be separated from the main components even for the shortest GRBs. 
The duration of the second component 
is $\tau_2 \simeq{\rm max}[\tau_{_\Omega},~\Delta\tau_f]$. 

The third component that may be observed in the light curve of a GRB
is generated at the distance $r\simeq r_{\rm dec}\sim 
10^{17}$~cm where the outflowing
wind with a helical magnetic field decelerates due to 
its interaction with an interstellar gas. The time delay between the 
precursor and  the third component is $\Delta\tau _{\rm dec}\simeq 
r_{\rm dec}/2c\Gamma_0^2$. Numerically, we have that 
$\Delta\tau _{\rm dec}$ is $\sim 10$~s at $\Gamma_0\simeq 300$ and 
$\sim 100$~s at $\Gamma_0\simeq 100$. 

The GRB light curves expected in our model depend on both the angle 
$\vartheta$ between the rotational and magnetic axes and the angle $\chi$
between the rotational axis and the sight light.
If $\vartheta$ is about
$\pi /2$, almost all the energy of the Poynting flux - dominated wind 
is radiated at the distance $r_f\sim 10^{13}-10^{14}$~cm
and transferred to the second component. 
In this case, the third component is strongly suppressed. 
On the contrary, if the rotational and magnetic axes are nearly aligned, 
the third component strongly dominates in GRB light curves
irrespective of the $\chi$ value. In this case, only a weak second component 
may be observed.
Such a weak second 
component may imitate a precursor radiated from the wind photosphere, but
its spectrum should be nonthermal. 
Therefore, the nature of a precursor may be find out only when its 
spectrum is measured. 
In general case, when $\vartheta$ is somewhere between zero and
$\pi /2$, both the second and third components with comparable
intensities may be seen  in 
GRB light curves. 


 In our model,  the time delay between the 
precursor and the second component cannot be much  more than 
the rise time of the second component. Figure~2a shows the light curve
of GRB 000508a with a weak precursor that is a plausible candidate to be 
radiated from the wind photosphere. For this GRB, the bulk of its emission 
is, most probably, the second component radiated at $r\sim r_f$.  
In the light curves of many GRBs (GRB970411, GRB910522, GRB991216, 
GRB 000508a, etc.), weak precursors separated by long ($\sim 10-100$~s)
background intervals were observed (see Fig.~2b). In this case, the
bulk of the GRB emission is naturally to identify with the third
component. However, to be sure that a weak first episode in GRB light 
curves is a precursor radiated from the wind photosphere,
it is necessary to make sure that its spectrum is nearly blackbody.
This may be done for strongest GRBs in near future with GLAST.

\section{Discussion}

Time histories of long GRBs often show multiple episodes of emission with
the count rate dropping to the background level between adjacent episodes
(e.g., Fishman et al. \cite{Fishman94}; Fishman \& Meegan \cite{FM95};
Meegan \cite{Meegan98}).
This is well consistent with the idea that there are several space-separated 
regions where the radiation of a GRB is produced. For some GRBs, the 
first episode is rather weak and may be treated as a precursor radiated 
from the wind photosphere. We expect that the most (if not all) GRBs 
have weak precursors with blackbody-like spectrum. The peak intensity
of a precursor is typically about $10-10^2$ times less than that of
the remaining GRB emission.

Koshut et al. (\cite{Koshut95}) estimated that $\sim 3$\% of the GRBs 
observed with BATSE exhibit precursor activity. 
However, their definition of precursor activity differs significantly
from ours. Indeed,
precursor activity was defined by Koshut et al. (\cite{Koshut95}) as 
any case in which the first episode has a lower peak intensity
than that of the remaining GRB emission and is separated from the 
remaining emission by a background interval that is at least as long 
as the remaining emission. In our model, one of the main observational 
features of a precursor radiated from the wind photosphere 
is its blackbody-like spectrum, and it is not necessary for 
the count rate to drop to a background level at all. Therefore, 
a weak first episode in the light curve of GRB000508a 
which is slightly ahead of the remaining emission (see Fig.~2a) is a 
reasonable candidate to be a precursor in our model 
while from the definition of Koshut et al. (\cite{Koshut95}) it follows 
that GRB000508 has no any precursor activity. However, this does not 
mean that the fraction of GRBs with already observed precursor 
activity in our conception is higher than $\sim 3$\%. The
reason is that for the main part of GRBs discussed in (Koshut et al. 
\cite{Koshut95}) the first episodes have a peak 
intensity either comparable or only a few times smaller than that of
the remaining emission. Such a first episode  is, most 
probably, the second components generated at $r\sim r_f$ 
while the remaining emission is the third components generated at
$r\sim r_{\rm dec}$.

\begin{acknowledgements} 
We  thank Norm Murray for useful comments.
V.V.U. is grateful for hospitality of the CITA  
where some of this work was carried out.
This work was supported in part by the MINERVA Foundation, Munich, Germany.
\end{acknowledgements}

\newpage

\begin{figure}[h]
\psfig{file=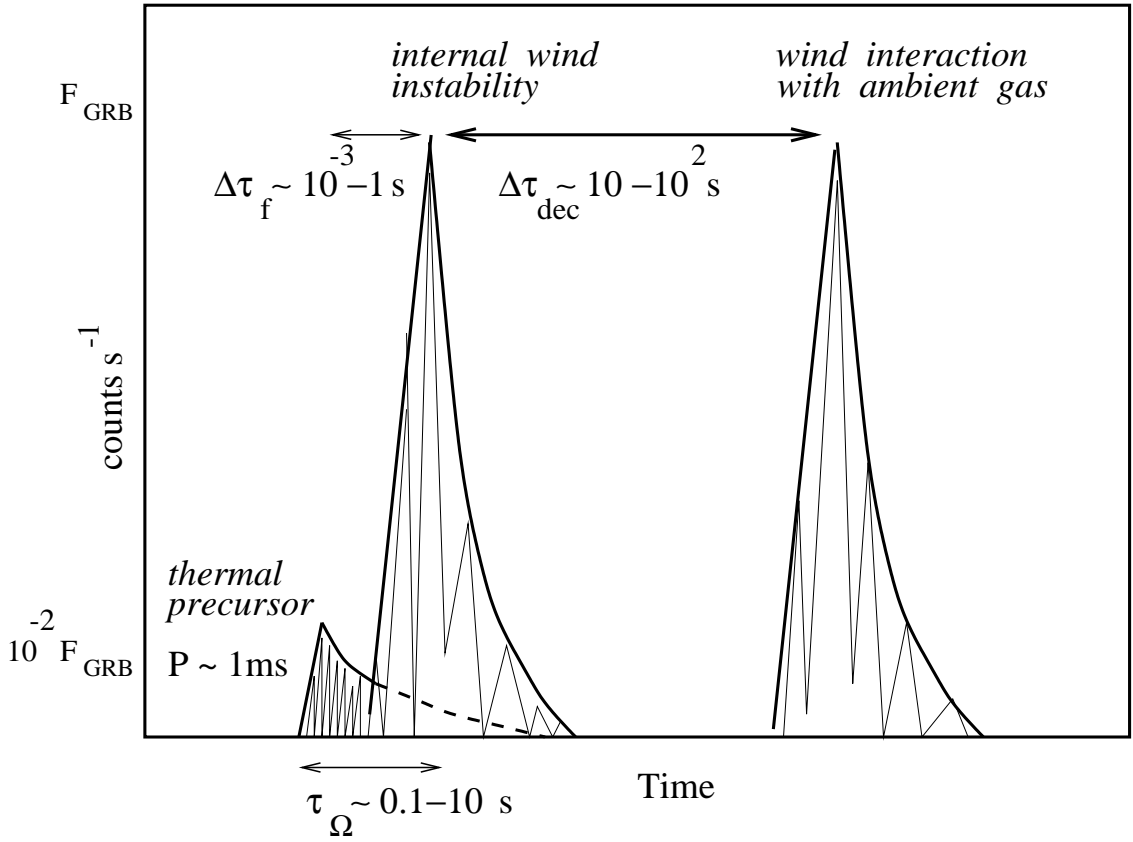,width=15cm}
\caption{Sketch of the light curves expected for GRBs.}
\label{prof}
\end{figure}

\newpage

\begin{figure}[h]
\begin{minipage}{0.45\textwidth}
\psfig{file=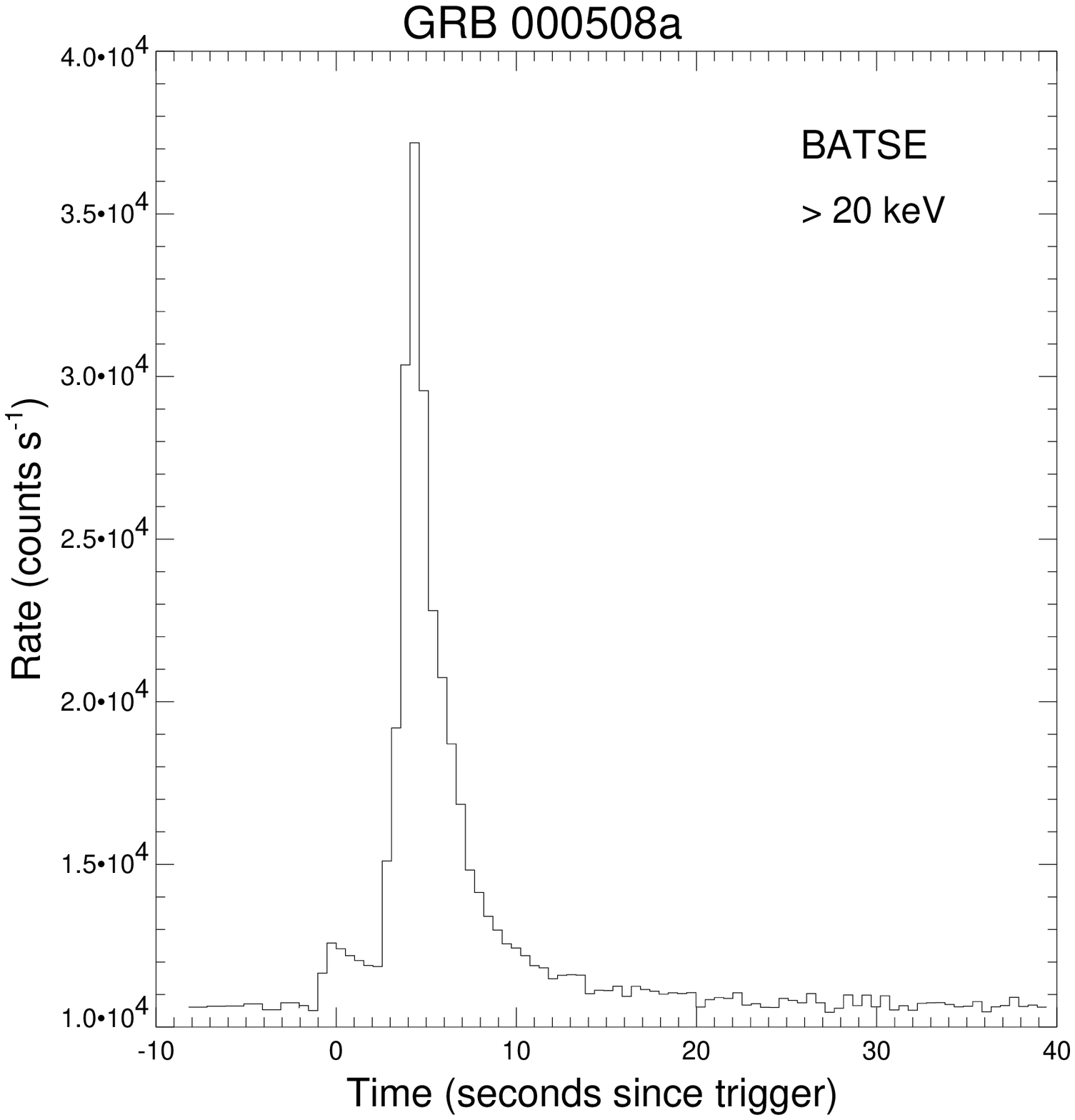,width=\linewidth}
\end{minipage}
\end{figure}
\begin{figure}[h]
\begin{minipage}{0.45\textwidth}
\psfig{file=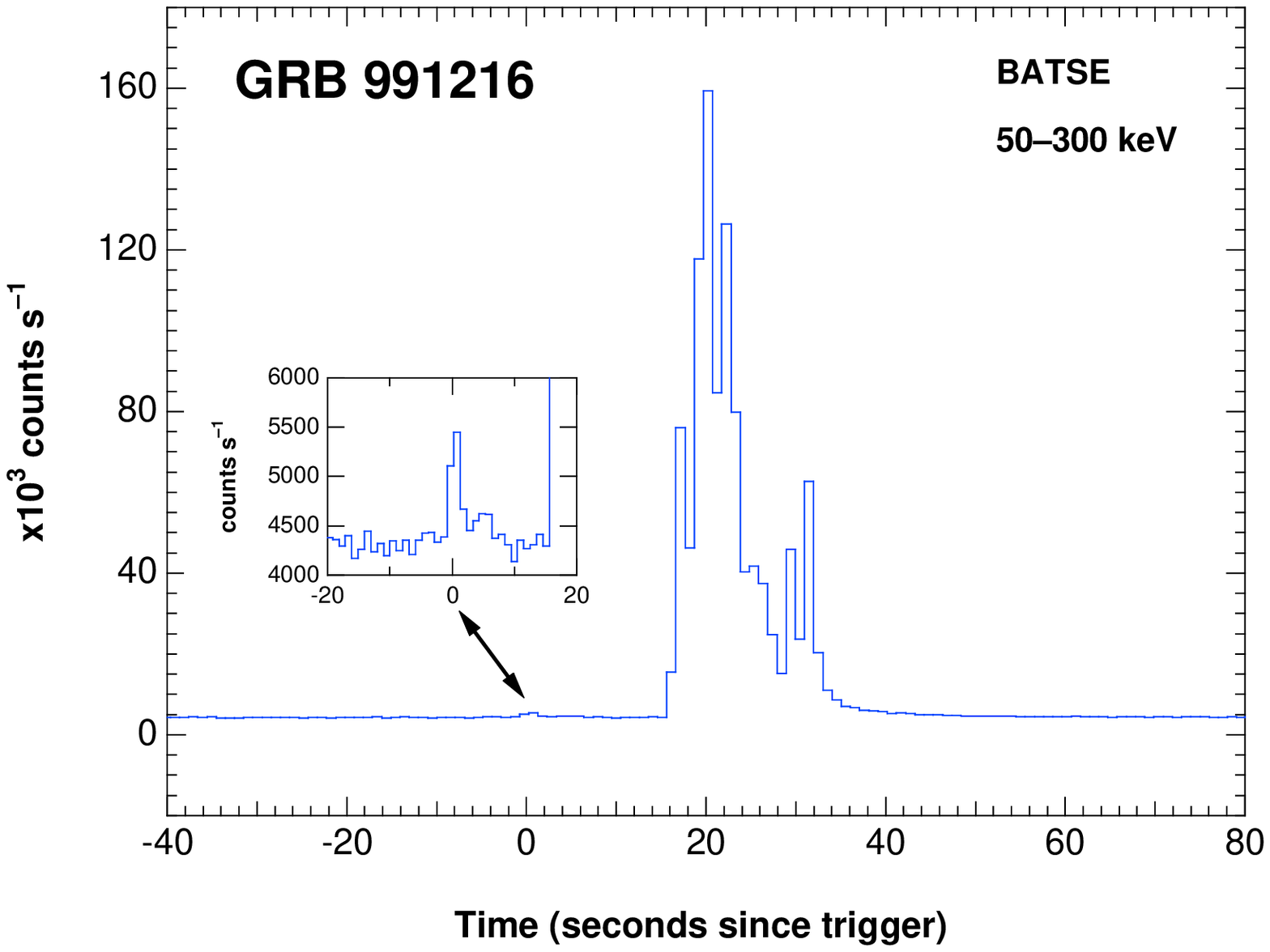,width=\linewidth}
\end{minipage}
\caption{Examples of GRBs where precursor may have been observed:
(a) GRB 000508, (b) GRB 991216.}
\end{figure}

\end{document}